\documentclass[12pt]{iopart}
\begin{document}

\title[Coupled Yang-Mills Oscillators I]
      {Adventures of the Coupled Yang-Mills Oscillators:\\
       I. Semiclassical Expansion}

\author{Sergei G.~Matinyan$^1$ 
\footnote[1]{Present address: 3106 Hornbuckle Place, Durham, NC 27707, USA.}
and Berndt M\"uller$^2$}

\address{$^1$ Yerevan Physics Institute, 375036 Yerevan, Armenia}

\address{$^2$ Department of Physics, Duke University, Durham, NC 27708}

\date{\today}

\begin{abstract}
We study the quantum mechanical motion in the $x^2y^2$ potentials
with $n=2,3$, which arise in the spatially homogeneous limit of the
Yang-Mills (YM) equations. These systems show strong stochasticity
in the classical limit ($\hbar = 0$) and exhibit a quantum mechanical 
confinement feature. We calculate the partition function $Z(t)$ 
going beyond the Thomas-Fermi (TF) approximation by means of the
semiclassical expansion using the Wigner-Kirkwood (WK) method. We derive 
a novel compact form of the differential equation for the WK function. 
After separating 
the motion in the channels of the equipotential surface from the motion 
in the central region, we show that the leading higher-order corrections 
to the TF term vanish up to eighth order in $\hbar$, if we treat the 
quantum motion in the hyperbolic channels correctly by adiabatic separation
of the degrees of freedom. Finally, we obtain an asymptotic expansion of 
the partition function in terms of the parameter $g^2\hbar^4t^3$.
\end{abstract}

%\pacs{}
\maketitle

\section{Introduction}

The discovery of the chaoticity of the classical Yang-Mills (YM) equations 
\cite{bib1} (see \cite{bib2} for a review) has attracted broad interest to 
the system of two (three) coupled quartic oscillators with the potential 
$x^2y^2$ ($x^2y^2+y^2z^2+z^2x^2$), where $x,y\,(z)$ are functions of time 
$t$. These systems are the simplest limiting cases for the homogeneous 
YM equations (sometimes called YM classical mechanics) with $n=2$ ($n=3$) 
degrees of freedom, respectively.The $x^2y^2$ model, the central object 
of the present paper, exhibits a rich chaotic behavior despite its extreme 
simplicity. Not surprisingly, this model has been encountered 
in various fields of science, including chemistry, astronomy, 
astrophysics, and cosmology (chaotic inflation). 

Quantum mechanically, this model (YM quantum mechanics, YMQM), despite 
possessing a logarithmically divergent volume of energetically 
accessible phase space \cite{bib2,bib3} (we set $m=1$ throughout)
\footnote{This is in violation of Weil's famous theorem, which states 
that the average number $N(E)$ of energy levels with energy less than 
$E$ is asymptotically proportional to $\int_0^E\Gamma_{E'} dE'$.}
\begin{equation}
\Gamma_E = \int_{-\infty}^{\infty} dx dy dp_x dp_y\, \delta\left(  
  \frac{1}{2}(p_x^2 + p_y^2) + \frac{g^2}{2} x^2 y^2 - E\right) ,
\label{eq00}
\end{equation}
can be shown to have a discrete spectrum \cite{bib5,bib6}. 
Physically, it is clear why 
this is so: Quantum fluctuations, e.g.\ zero-point fluctuations, forbid 
the ``particle'' to escape along the $x$ or $y$ axis where the potential 
energy vanishes. The system is thus confined to a finite volume, and 
this implies the discreteness of the energy levels. Classically, of 
course, the particle can always escape along one of the axes without
increasing its energy. 

In this article we calculate the partition function (heat kernel) $Z(t)$ 
for the YMQM with the above given potentials beyond the well-known 
Thomas-Fermi (TF) approximation, which takes into account only the 
discreteness of the quantum mechanical phase space, but treats the 
Hamiltonian classically. There are several interesting articles
\cite{bib7,bib8} devoted to the approximate calculation of $Z(t)$ in the 
TF approximation for these potentials. They are based on the adiabatic 
separation of the dependence of $Z$ on $x$ and $y$ in the narrow channels 
of the equipotential surface $xy = {\rm const.}$. The range of the 
integration over the coordinates $x,y$ and the momenta $p_x,p_y$ was 
divided into two regions: the central region ($|x|,|y| \leq Q$) and 
the channels ($Q \leq |x|, Q \leq |y|$), which are governed by quite different 
physics: The system is essentially classical in the central region and 
intrinsically quantum mechanical in the channels performing oscillatory motion 
with $x$-dependent frequency in the ``fast'' variable $y$ (in the channels 
along the $x$-axis), but quasi-free motion in the ``slow'' variable $x$.

The dependence on the artificial boundary Q dividing the 
central region from the channels disappears in the final answer for $Z(t)$.
Below we show that this property survives beyond the TF approximation up
to the order $\hbar^8$. In the paper \cite{bib9} an alternative to this 
approach of calculating $Z(t)$ was proposed, starting from the 
Yang-Mills-Higgs quantum mechanics (YMHQM) and then passing to the limit 
$v =0$, where $v$ is the vacuum expectation value of the ``Higgs field'' 
defining the strength of the harmonic potential. 

Here we study the $x^2y^2$ potential and postpone the investigation of YMHQM 
to a separate publication \cite{bib*}. We here follow the method of separation
of the domain of motion into two regions, the central square $x,y\in
[-Q,Q]$ and the hyperbolic channels, introduced by Tomsovic \cite{bib7} 
and Whelan \cite{bib8}. The motion in the channels can be treated by
adiabatic separation of the motion in the slow variable (in the direction 
of the channel) and the motion in the fast variable (perpendicular to
the channel). We will go beyond the Thomas-Fermi approximation used
in \cite{bib7,bib8} by applying the well-known Wigner-Kirkwood (WK)
method \cite{bib10,bib11,bib12} (see \cite{bib13} for a review of the 
WK approach). 

We already mentioned that the motion in the central region, where the
variables $x$ and $y$ are treated on an equal footing, is quite different
from the motion in the channels, where the motion in the perpendicular
direction (for now taken as the variable $y$) performs quantum oscillations 
with $x$-dependent frequency, but the motion in the $x$ variable is not
treated as free as it was done in \cite{bib7,bib8}. We apply the WK
method in the central region to both variables, but only to the slow
variable $x$ in the channels treating the motion in the fast $y$ variable
fully quantum mechanically. We improve the method of adiabatic separation
introduced in \cite{bib7,bib8} by taking into account the effect of
the $x$-dependence of the oscillation frequency onto the motion in the
$x$-direction. This is especially important, as we shall see, for the
higher-order quantum corrections. We show that, up to eighth-order in
$\hbar$ and in the leading terms in $(tQ^4)^{-1} \ll 1$, the calculated
partition function $Z(t)$ does not depend on the boundary $Q$ dividing
the two regions and the quantum corrections from both regions cancel.

An interesting phenomenon occurs in the channel motion due to the presence
of terms suppressed by powers of $(tQ^4)^{-1}$, which are independent
of $Q$. Each term of this kind is negligible in comparison with the
leading (but canceling) terms, but their series forms an asymptotic
series in the variable $g^2\hbar^4t^3$, which does not involve $Q$.
If we postulate that the entire $Q$-dependence of the partition function
disappears, which we prove for the leading terms in $(tQ^4)^{-1}$,
only this series will be left as the contribution from the channels.
We may express this phenomenon as the ``transmutation'' of the small 
expansion parameter $(tQ^4)^{-1}$ governing the approach of adiabatic
separation of variables into the small parameter $g^2\hbar^4t^3$ 
characteristic of quantum mechanics. The quasiclassical expansion is 
shown to have the nature of an asymptotic series.

In the next two sections we present the YMQM system and the WK method of 
calculating $Z(t)$ beyond the TF approximation.

\section{Yang-Mills-Higgs classical and quantum mechanics: The
         Thomas-Fermi approximation}
 
For spatially homogeneous fields (long wave length limit of the Yang-Mills
field) the classical Hamiltonian for $n=2$ is given by the expression 
\begin{equation}
H = \frac{1}{2}(p_x^2+p_y^2) + \frac{g^2}{2}x^2y^2 .
\label{eq01}
\end{equation}
The quantized counterpart of (\ref{eq01}) is 
\begin{equation}
\hat H = -\frac{\hbar^2}{2}\nabla_{x,y}^2 + \frac{g^2}{2}x^2y^2 .
\label{eq02}
\end{equation}
A brief word on units: All quantities are given below in units of 
the energy $E$ with dimensions $[H]=1, [t]=-1, [x],[y]=1/4, [g]=0, 
[\hbar]=3/4$. The operator (\ref{eq02}) has a discrete spectrum 
\cite{bib5,bib6}. The TF approximation to the heat kernel 
or partition function $Z(t) = {\rm Tr} [\exp(-t\hat H)]$ 
is the standard lowest-order semiclassical approximation valid for 
small $\hbar t^{3/4} \ll 1$. It is obtained by substituting the 
classical Hamiltonian for its quantum counterpart and replacing the 
trace of the heat kernel by the integral over the phase-space volume 
normalized by ($2\pi\hbar)^{-n}$, where $2n$ is the phase-space 
dimension. In other words, the TF approximation takes into account 
only the discreteness of the quantum mechanical phase space, but 
considers momenta and coordinates (in our case, the field amplitudes 
$x$ and $y$) as commuting variables. This method was used in numerous 
papers (see e.g. \cite{bib7,bib8,bib9}). For the calculation of the 
energy level density $\rho(E) = dN(E)/dE$ at asymptotic energies, the TF 
approximation is a consistent approach since, as we shall see below, all 
corrections to the TF term are structures with factors $\hbar^k t^\ell$ 
with $k,\ell$ positive integers. For the asymptotic energy level density 
$\rho(E)$ or $N(E)$ these corrections are negligible according the 
Karamata-Tauberian theorems \cite{bib5,bib6} relating the most singular 
part of $Z(t)$ to the asymptotic level density: 
$N(E) = \int dE \rho(E)= L^{-1}(Z(t)/t)$ where $L^{-1}$ denotes the 
inverse Laplace transform. 

For the Hamiltonian (\ref{eq02}) the naive TF approximation is divergent, 
because the classical phase space is infinite. As explaines in the
Introduction, finite results for $Z(t)$  and $N(E)$ can be obtained by 
including certain quantum corrections to the channel motion by means of 
the method of adiabatic separation of the motion along and perpendicular 
to the channels \cite{bib7,bib8}. We here give the expression for
the partition function obtained in this improved TF approximation in
our notation:
\begin{equation}
Z_0 = \frac{1}{\sqrt{2\pi}g\hbar^2 t^{3/2}} 
       \left(\ln\frac{1}{g^2\hbar^4t^3} + 9\ln 2 + C\right) ,
\label{eq04}
\end{equation}
where $C$ is the Euler constant. We shall return to (\ref{eq04}) 
again below. Because we shall often encounter the pre-factor appearing 
in (\ref{eq04}), we introduce the special symbol for it:
\begin{equation}
K \equiv (2\pi g^2\hbar^4 t^3)^{-1/2} .
\label{eq04a}
\end{equation}

\section{Beyond the TF approximation: The Wigner-Kirkwood expansion}

For non asymptotic energies and also for the fluctuating part of the 
level density one needs to go beyond the TF approximation and calculate 
the quantum corrections to the TF term in $Z(t)$. We remark that this 
problem is interesting not only from this practical point of view, but also
because it provides insight into some problems of perturbation theory 
in quantum mechanics. Here we use the Wigner-Kirkwood (WK) method. 
Before we proceed, we emphasize the dual role of the variables 
$x(t),y(t),z(t)$. They formally play the role of the coordinates, but
at the same time they stand for the homogeneous gauge field amplitudes. 
The trace in the heat kernel may be taken with respect to any complete 
set of states. For the semiclassical expansion involving integration over 
the phase space for the quantum operators in the Wigner representation 
it is convenient to use the plane waves as a complete set: 
\begin{equation}
Z(t) = \frac{1}{(2\pi\hbar)^n} \int \prod_{i=1}^{n} dx_i dp_i \,
       e^{-i{\vec p}{\vec r}/\hbar} e^{-t{\hat H}} e^{i{\vec p}{\vec r}/\hbar}
\label{eq05}
\end{equation}
with $\vec r = (x_1,x_2,\ldots,x_n), \vec p = (p_1,p_2,\ldots,p_n)$. 
The kinetic energy term in the exponent of $\exp(-t{\hat H})$ does not 
commute with the potential energy $V({\vec x})$; the WK expansion
provides a convenient method of calculating the noncommuting terms. 
Following \cite{bib12} we set
\begin{equation}
e^{-t{\hat H}} e^{i{\vec p}{\vec r}/\hbar} 
= e^{-tH} e^{i{\vec p}{\vec r}/\hbar} W({\vec r},{\vec p};t)
= u({\vec r},{\vec p};t)
\label{eq06}
\end{equation}
where the function $W({\vec r},{\vec p};t)$ is to be determined.
$H({\vec p},{\vec r})$ (as opposed to $\hat H$) is the classical Hamiltonian 
(\ref{eq01}). The function $u({\vec r},{\vec p};t)$  satisfies the 
Bloch equation: 
\begin{equation}
\frac{\partial u}{\partial t} + {\hat H} u = 0 
\label{eq07}
\end{equation}
with the boundary condition 
\begin{equation}
\lim_{t\to 0} u({\vec r},{\vec p};t) = e^{i{\vec p}{\vec r}/\hbar},
\label{eq08}
\end{equation}
corresponding to the initial condition $W({\vec r},{\vec p};0) = 1$.
From (\ref{eq06}) and (\ref{eq07}) we obtain an exact equation for $W$: 
\begin{eqnarray}
\frac{\partial W}{\partial t} 
&=& \frac{\hbar^2}{2}\left[ \Delta - t(\Delta V) 
  - \frac{2it}{\hbar}({\vec p}\cdot\nabla V) + t^2(\nabla V)^2 \right.
\nonumber \\
&&\qquad  \left. 
  + \frac{2}{\hbar}(i{\vec p} 
  - \hbar t\nabla V)\cdot\nabla\right] W ,
\label{eq09}
\end{eqnarray}
where $\Delta = \nabla^2$ is the Laplacian. Next we expand $W$ in 
powers of $\hbar$:
\begin{equation}
W = \sum_{k=0}^{\infty} \hbar^k W_k  
\label{eq09b}
\end{equation} 
and equate the terms with the same power of $\hbar$ on both sides. We 
thus obtain a recurrence relation of differential equations for the
$W_k$:
\begin{eqnarray}
\frac{\partial W_k}{\partial t} 
&=& \frac{1}{2}\left[ \Delta - t(\Delta V) + t^2(\nabla V)^2 
  - 2t\nabla V\cdot\nabla\right] W_{k-2} 
\nonumber \\
&&  + i{\vec p}\cdot\left[\nabla - t(\nabla V)\right] W_{k-1} ,
\label{eq10}
\end{eqnarray}
with the initial conditions $W_k=0$ for $k<0, W_0=1$. Since $Z(t)$ and $W$ 
are linearly related, the expansion of $W$ in powers of $\hbar$ immediately
translates into an expansion of $Z(t)$ in powers of $\hbar$. 
To obtain the term $Z_k(t)$, one needs to calculate $W_k$ from (\ref{eq10}) 
and integrate over $\vec p$ and $\vec x$: 
\begin{equation}
Z_k(t) = \frac{\hbar^k}{(2\pi\hbar)^n} \int_{-\infty}^{\infty}
  \prod_{i=1}^n dx_i dp_i W_k({\vec r},{\vec p};t)
  \exp\left[-t\left(\frac{{\vec p \,}^2}{2} + V(\vec x)\right)\right] .
\label{eq11}
\end{equation}

The expressions (\ref{eq09}) and (\ref{eq10}) may be written in more 
compact form, if we introduce the ``covariant derivative''
${\vec D}\equiv \nabla - t\nabla V$:
\begin{equation}
\frac{\partial W}{\partial t} = \frac{\hbar^2}{2}
  \left[ D^2 + \frac{2i}{\hbar}{\vec p}\cdot{\vec D} \right] W 
= \frac{1}{2} \left[ (\hbar{\vec D}+i{\vec p}\,)^2 + {\vec p}\,^2 \right] W
\label{eq11a}
\end{equation}
or its  recursive form
\begin{equation}
\frac{\partial W_k}{\partial t} = \frac{1}{2} \left[ D^2 W_{k-2} 
  + 2i{\vec p}\cdot{\vec D}W_{k-1} \right] .
\label{eq11b}
\end{equation}
We emphasize that the symbol $\vec p$ here denotes a classical phase-space
variable and not an operator. As far as we know, the form (\ref{eq11a}) of 
the equation (\ref{eq09}) has not previously been presented in the literature.

In terms of the operator $\vec D$, eq.~(\ref{eq11a}) resembles a Fokker-Planck
equation for $W({\vec r},{\vec p};t)$ with the diffusion constant 
$\sigma=\hbar^2$ and the drift vector ${\vec\gamma}=-i\hbar\vec p$ (see text 
below eq.~(\ref{eq11c})). The relation to the Fokker-Planck equation
can be further elucidated by noting that the ``vector potential'' 
${\vec A}=t\nabla V$ is a complete gradient and thus can be ``gauged'' away 
by means of the transformation $W \to e^{tV}W'$, yielding an alternative 
form of (\ref{eq11a}):
\begin{equation}
\frac{\partial W'}{\partial t} = \frac{\hbar^2}{2}
  \left[ \nabla^2 + \frac{2i}{\hbar}{\vec p}\cdot\nabla \right] W'
  - V W' .
\label{eq11c}
\end{equation}
If we interpret $W'({\vec r},{\vec p};t)$ as a one-time probability density
and introduce the probability current (sometimes called the probability 
flux in the literature) \cite{bibH}
\begin{equation}
{\vec J}\,' = -i\hbar {\vec p}\, W' - \frac{\hbar^2}{2}\nabla W' ,
\label{eq11c1}
\end{equation}
we may write (\ref{eq11c}) in the form of a continuity equation:
\begin{equation}
\frac{\partial W'}{\partial t} + \nabla\cdot{\vec J}\,' = - V W' ,
\label{eq11d}
\end{equation}
where the potential term acts as a source term and violates the local 
conservation law associated with the Fokker-Planck equation.

The recurrence relation (\ref{eq11b}) clearly shows that the expansion in 
$\hbar$ introduced by Kirkwood \cite{bib11} is also an expansion in powers 
of the gradient operator as emphasized by Uhlenbeck and Beth \cite{bib12}. 
In the general case one needs to expand in powers of $\hbar$ or, 
equivalently, in powers of the gradient operator using (\ref{eq10})
or (\ref{eq11b}), as we will do here. In special cases, however, the
compact form of the equations (\ref{eq11a}) or (\ref{eq11c}) may be the 
starting point of other effective approximation schemes.

\section{Testing the method of separation beyond the TF 
         approximation for the $x^2y^2$ potential}

We now want to check whether the method of the separation 
$x$- and $y$- motions in the channels \cite{bib7,bib8} works 
beyond the TF approximation. In the TF approximation, where the
integrand in $Z(t)$ is simply $\exp(-tH)$, the boundary $Q$ appears in the
argument of a logarithm. Higher WK corrections to the TF term lead to a 
power-like dependence on $Q$, and it is important to confirm that all 
dependence on $Q$ is cancelled in the final answer, at least for the
leading terms in the parameter $tQ^4 \gg 1$.  Here we consider this 
problem up to the second-order corrections. In the later sections 
we will consider corrections up to the order $\hbar^8$.

As we stressed before, we apply the WK method to the motion in the
central region ($|x|,|y| \leq Q$) in its full scope, whereas for the
motion in the channels ($|x|,|y| \geq Q$) this method will be used only
to study the quantum effects on the ``slow'' longitudinal motion, which
is adiabatically separated from the quantized oscillatory motion in the
transverse coordinate.

Integrating (\ref{eq10}) with respect to $t$, we have:
\begin{eqnarray}
W_1 &=& -\frac{it}{2} {\vec p}\cdot\nabla V
\nonumber \\
W_2 &=& \frac{t^2}{2} \left[ -\frac{1}{2}\Delta V + \frac{t}{3}(\nabla V)^2
    +\frac{t}{3}\sum_{i,k} p_ip_k\frac{\partial^2V}{\partial x_i \partial x_k}
    -\frac{t^2}{4}({\vec p}\cdot\nabla V)^2 \right]
\label{eq12}
\end{eqnarray}
with $V(x,y)=\frac{1}{2}g^2x^2y^2$. Integration over $p_x,p_y$ and making
use of the symmetry of the Hamiltonian with respect to the interchange
$x \leftrightarrow y$, we find that the contribution from $W_1$ vanishes
and
\begin{equation}
\int d\Gamma\, W_2\, e^{-tV} 
= \frac{\pi tg^2}{3} \left[ -I_{10} + \frac{tg^2}{2} I_{21} \right] ,
\label{eq20}
\end{equation}
where we introduced the abbreviations
\begin{equation}
d\Gamma = dxdydp_xdp_y
\end{equation}
and (for $m\geq n$)\footnote{Note that the case $m=n$ needs to be calculated
separately from the case $m>n$ for the leading terms; see below.}:
\begin{equation}
I_{mn} = 4 \int_0^Q dx \int_0^Q dy\, x^{2m} y^{2n} e^{-tg^2x^2y^2/2} .
\label{eq21}
\end{equation}
These integrals can be evaluated straightforwardly after the substitution 
$x=w$ and $y=\sqrt{2/t}(u/gw)$ and with the help of the condition for 
the validity of the adiabatic approximation $Qt^{1/4}\gg 1$. Note that
this inequality does not contradict the condition permitting the use of
the Wigner representation $\hbar Qt \ll 1$ if $\hbar t^{3/4} \ll 1$. 
We obtain 
\begin{equation}
I_{10} \approx \frac{\sqrt{2\pi}}{gt^{1/2}} Q^2 , \qquad
I_{21} \approx \frac{\sqrt{2\pi}}{g^3t^{3/2}} Q^2 .
\label{eq21a}
\end{equation}
and finally for $Z_2(t)$ in the square:
\begin{equation}
Z_2 = - K \frac{1}{12} (g\hbar tQ)^2 . 
\label{eq22}
\end{equation}
For the sake of completeness and later use, we note the systematic
structure of the integrals $I_{mn}$. For a given value of $m-n$
there are $m-n+1$ such expressions: $I_{m-n,0},I_{m-n+1,1},\ldots ,
I_{2(m-n),m-n}$. Applying the same method as above yields the following
result for these expressions at fixed $m-n$:
\begin{equation}
I_{mn} = \frac{\sqrt{2\pi}}{(gt^{1/2})^{2n+1}}\frac{(2n-1)!!}{m-n}
          Q^{2(m-n)} .
\label{eq22b}
\end{equation}

For the second region (the four channels $|x|\in [Q,\infty]$ or 
$|y|\in [Q,\infty]$), the integration is more involved. 
WK corrections to the TF term introduce 
pre-exponential functions of $p_x,p_y,x,y$ and $t$ in $W_2$. Because 
of the fourfold symmetry of the Hamiltonian (\ref{eq02}) it is sufficient
to consider the channel $x\ge Q$. Closely following the method used in 
\cite{bib7,bib8} we consider the motion in the $x$ variable as free and the 
motion in the $y$ variable as that of a harmonic oscillator with an 
$x$-dependent frequency with the Hamiltonian 
\begin{equation}
H_y = \frac{1}{2}p_y^2 + \frac{1}{2}\omega_x^2 y^2 
\label{eq22a}
\end{equation}
with $\omega_x=gx$ and eigenvalues $e_n(x) = (n + \frac{1}{2}) \hbar gx$.

In order to reduce the integrals over $p_y^2$ and $y^2$ to derivatives
of the well-known exponentiated sum-rule for the harmonic oscillator, we 
rescale the kinetic energy term in $H_y$ by a factor $b$ and the
potential energy term by a factor $a$. We denote the rescaled Hamiltonians
by $H'_y$. We then can generate any pre-exponential powers of 
$p_y^2$ and $y^2$ by differentiating with respect 
to $a$ and $b$ and setting $a=b=1$ in the final result. 
Integrating $W_2$ from (\ref{eq12}) over $p_x$ and $x$ and performing 
the described substitutions and differentiations we obtain :
\begin{equation}
\int_{-\infty}^{\infty} dp_x \int_Q^{\infty} dx\, W_2\, 
     {\rm Tr}\left(e^{-tH'_y}\right)
=  2 \sqrt{\frac{2\pi}{t}} (gt)^2 \int_Q^{\infty} dx D(x)
\label{eq23}
\end{equation}
with
\begin{eqnarray}
D(x) &=& \left[ - \frac{x^2}{2} - \frac{2x^2}{3} \frac{d}{da} 
     + \frac{1}{3g^2tx^2} \frac{d^2}{da^2} \right.
\nonumber \\
   && \left.  \qquad\qquad\qquad
     - \frac{2x^2}{3} \frac{d}{db} 
     + x^2 \frac{d^2}{dadb} \right]
      {\rm Tr}\left(e^{-tH'_y}\right)_{a=b=1} .
\label{eq23p}
\end{eqnarray}
In (\ref{eq23}) we may discard the third term as it is 
${\cal O}(1/x^4t) \ll 1$ with respect to other terms ($x>Q,Q^4t\gg 1$ in 
the channel).\footnote{One may notice that contributions arising from the 
derivative with respect to the ``fast'' variable $y$ in $W_2$ are not 
small, whereas derivatives with respect to $x$ are kinematically 
negligible in the channel where $Q^4t\gg 1$.} 
We note that the algebraic trick dealing with terms including the ``fast'' 
variables $y,p_y$ effectively reduce the problem in the channels to the 
level of TF terms with the rescaled Hamiltonians $H'_y$. 
Now we can use the exponentiated sum-rule for the harmonic oscillator 
(see \cite{bib7}) 
to obtain 
\begin{eqnarray}
D(x) &=& \left[ -\frac{x^4}{4\sinh \xi}
     - \frac{2x^2}{3} \frac{d}{da}\frac{1}{\sinh(\xi\sqrt{a})} \right.
\nonumber \\
   && \left. \qquad\qquad\qquad
     - \frac{x^2}{2} \frac{d^2}{dadb}\frac{1}{\sinh(\xi\sqrt{ab})}
       \right]_{a=b=1} .
\label{eq24}
\end{eqnarray}
where $\xi = \hbar gtx/2$. 
Performing the differentiations, setting $a=b=1$, and multiplying by 4
to account for all four channels, we obtain for the channel contribution
to $Z_2(t)$:
\begin{eqnarray}
Z_2(t) &= 4 K
   \int_{\xi_0}^\infty \xi^2 d\xi\, & \left[ -\frac{1}{\sinh\xi} 
      + \frac{11}{6} \xi \frac{\cosh\xi}{\sinh^2\xi} \right.
\nonumber \\
   && \left. \qquad 
      + \frac{\xi^2}{2\sinh\xi} -\frac{\xi^2\cosh\xi}{\sinh^3\xi} \right] ,
\label{eq25}
\end{eqnarray}
with $\xi_0=\hbar gtQ/2 (\ll 1)$. Performing the integrations in (\ref{eq25}) 
(see \cite{bib17}, integrals 2.477.3, 3.523.1, 2.479.4, 2.477.1, 3.523.1), 
we obtain for the channel contribution to $Z_2$: 
\begin{equation}
Z_2(t) = K \left[ \frac{1}{12}(g\hbar tQ)^2 - 21 \zeta(3) \right] .
\label{eq26}
\end{equation}
where $\zeta(z)$ is the Riemann zeta function ($\zeta(3)\approx 1.202$). 
From (\ref{eq22}) and (\ref{eq26}) one sees that the $Q$-dependence of 
$Z_2(t)$ cancels, and the final answer is 
\begin{equation}
Z_2 (t) \approx -25.2 K .
\label{eq27}
\end{equation}
We achieved the $Q$-independence of the second-order correction to the 
partition function of $x^2y^2$ model using the method of \cite{bib7,bib8}.
The huge renormalization of the TF term in (\ref{eq27}) looks suspicious.

\section{Improved quantum motion in the channels}

The unexpectedly large coefficient in (\ref{eq26}) suggests that we need 
to find a better treatment for the motion in the channel.
The previous study of the quantum motion in the channel 
neglected the quantum nature of the ``slow'' motion along the channel axis, 
which was assumed to be free in \cite{bib7,bib8}. We already stressed in the 
Introduction that this motion is by no means free, rather, it is influenced 
by an effective linear potential created by the quantum fluctuations in the
direction(s) orthogonal to the channel axis. Indeed, in the region 
$|x|\gg |y|$, where the derivatives with respect to $x$ are small relative 
to the ones with respect to $y$, we may first average the motion over the 
quantum fluctuations of $y$ \cite{bib18} described by the Hamiltonian $H_y$ 
(\ref{eq22a}) and by the corresponding wave function 
\begin{equation}
\psi_n(y) = \frac{1}{\sqrt{2^n n!}} \left(\frac{gx}{\pi\hbar}\right)^{1/4}
  e^{-gxy^2/2\hbar} H_n(y\sqrt{gx/\hbar}) ,
\label{eq28a}
\end{equation}
where $H_n(z)$ are the Hermite polynomials. The corresponding average value 
of $H_y$ 
\begin{equation}
\langle n| H_y |n \rangle = (n+\frac{1}{2})\hbar gx 
\label{eq29}
\end{equation}
then becomes an effective potential for the description of the motion 
in the ``slow'' variable $x$: 
\begin{equation}
\left( -\frac{\hbar^2}{2}\frac{\partial^2}{\partial x^2}
  + (n+\frac{1}{2})\hbar gx \right) \phi_n(x) = E \phi_n(x) .
\label{eq30}
\end{equation}
This is the well-known Schr\"odinger equation for a linear potential with 
solutions in terms of Airy functions. Equation (\ref{eq30}) clearly shows
that, quantum mechanically, the ``particle'' is linearly confined along the 
channel axis and its motion is not free.\footnote{Note that the phenomenon
called ``confinement'' here is not the same as the phenomenon commonly
referred to as quark confinement. In our case, the potential depends linearly
on the field amplitude $x(t)$, not on a spatial coordinate. We already
emphasized the dual role of the coordinates $x,y,z$ earlier, but we stress
this point again here to avoid misunderstandings. One may also have called
the phenomenon discussed here ``self-confinement'', as the fields themselves
``prepare'' the effective potential barrier prohibiting the escape to
infinity.} The eigenvalue problem (\ref{eq30}) has a 
discrete spectrum. Note that this argument constitutes a sixth proof, 
in addition to the five proofs listed in \cite{bib5}, for the discreteness
of the spectrum of the Hamiltonian (\ref{eq02}). It explains the large 
number in (\ref{eq26}) and (\ref{eq27}) as an artefact of the assumption 
of free motion in the $x$-direction. This assumption becomes increasingly 
poor for the higher-order quantum corrections, leading to poor convergence 
or even divergence of the expansion in powers of $\hbar$. Treating the 
``slow'' $x$-motion in the channel adiabatically, we apply the WK expansion 
and eq.~(\ref{eq10}) to the effective Hamiltonian 
\begin{equation}
H^{(n)}_x = \frac{1}{2}p_x^2 + (n+\frac{1}{2})\hbar gx 
\label{eq31}
\end{equation}
describing the motion of the $n$th quantum mode in the region $x\ge Q$. 
For the partition function of the one-dimensional motion of the $n$th mode 
we have (denoting $p_x$ simply by $p$): 
\begin{equation}
Z^{(n)}(t) = \frac{1}{2\pi\hbar} \int_{-\infty}^{\infty} dp 
\int_Q^\infty dx\, e^{-\frac{1}{2}p^2 t -(n+\frac{1}{2})\hbar gxt} W(x,p;t) .
\label{eq32}
\end{equation}
Expanding $W(x,p;t)$ in powers of $\hbar$ for the linear potential in 
(\ref{eq31}) and using eq.~(\ref{eq10}), we easily obtain: 
\begin{eqnarray}
\label{eq33}
W_2 &=& \frac{1}{3\cdot 2^3} a_n^2 t^3 (4 - 3p^2t) ;
\\
W_4 &=& \frac{1}{3^2\cdot 2^7} a_n^4 t^6 (16 - 24p^2t + 3p^4t^2)
\nonumber \\
W_6 &=& \frac{1}{2^{10}\cdot 3^3\cdot 5!!} a_n^6 t^9 (320 - 720p^2t
        +180p^4t^2 - 9p^6t^3) ,
\nonumber \\
W_8 &=& \frac{1}{2^{15}\cdot 3^4\cdot 7!!} a_n^8 t^{12}
        (8960 -26880p^2t +10080p^4t^2 -1008p^6t^3 +27p^8t^4) ,
\nonumber
\end{eqnarray}
where $a_n=(n+\frac{1}{2})\hbar g$. The terms with odd indices contain odd 
powers of $p$ and give vanishing contributions after integration over $p$.
Carrying out the integration over $p$ in (\ref{eq32}), summing over 
$n$, and retaining all terms with derivatives up to order $\hbar^8$, we 
obtain for the contribution from a single channel:
\begin{eqnarray}
\label{eq34}
Z_{\rm ch}(t) 
     &= \int_Q^\infty \frac{dx}{\sqrt{2\pi t}\hbar}
       & \left( 1 + \frac{\hbar^2t}{24}\frac{\partial^2}{\partial x^2}
              + \frac{\hbar^4t^2}{1152}\frac{\partial^4}{\partial x^4}
         \right.
\\ && \qquad \left.
              + \frac{\hbar^6t^3}{82944}\frac{\partial^6}{\partial x^6}
              + \frac{\hbar^8t^4}{7962624}\frac{\partial^8}{\partial x^8}
       + \cdots \right) \frac{1}{2\sinh\xi} ,
\nonumber
\end{eqnarray}
where again $\xi = \hbar gxt/2$. The expansion parameter $\hbar^2t/Q^2\ll 1$ 
in (\ref{eq34}) is the squared ratio of the ``diffusion'' length mentioned 
in Section 3 and the separation scale $Q$. After performing the 
differentiations and integrating over $x$ we finally 
have for all four channels:
\begin{eqnarray}
Z(t) = K \left[ 4 \ln\coth\frac{\xi}{2} \right.
       &-& \left( \frac{\lambda^2}{2^3\cdot 3}\partial_\xi
         + \frac{\lambda^4}{2^9\cdot 3^2}\partial^3_\xi
         + \frac{\lambda^6}{2^{14}\cdot 3^4}\partial^5_\xi \right.
\nonumber \\
  &&\qquad \left. \left.
         + \frac{\lambda^8}{2^{21}\cdot 3^5}\partial^7_\xi 
       \cdots \right)\frac{1}{\sinh\xi} \right]_{\xi=u}
\label{eq35}
\end{eqnarray}
with $u=\hbar gtQ/2$ and $\lambda^2=g^2\hbar^4t^3$. This expression,
with the exception of the logarithm, can be written as a power series 
in $u$ by substituting the Bernouilli expansion of the hyperbolic
cosecans:
\begin{equation}
\frac{1}{\sinh\xi} = \frac{1}{\xi} - 2 \sum_{n=1}^{\infty}
   \frac{2^{2n-1}-1}{(2n)!} B_{2n} \xi^{2n-1} .
\label{eq35a}
\end{equation}
where $B_{2n}$ denotes the Bernoulli numbers ($B_2 = 1/6, B_4 = -1/30,
B_6 = 1/42, B_8 = -1/30, \ldots$). 
It is then easy to see that there are three kinds of terms in the 
series involving derivatives $\partial^{2n-1}_{\xi}$ in (\ref{eq35}):
\begin{enumerate}
\item
terms with a positive power of $u^2$; there is an infinite number of
such terms;
\item
terms with an inverse power of $u^2$; there is one such term at for each
derivative;
\item
terms independent of $u$, and thus independent of $Q$; again there is 
only one such term for each derivative. 
\end{enumerate}
The terms of type (i) are of the order $\lambda^{2n}(\hbar Qt)^{k-4n}$,
where $k$ denotes the overall power of $\hbar$. The leading term in
$Z(t)$ at the same power of $\hbar$ being of order $(\hbar Qt)^k$,
arising from the expansion of the logarithmic term in (\ref{eq35}),
the contribution from the series is suppressed by a factor 
$(tQ^4)^{-n} \ll 1$. For the terms of type (ii) one can show that they
are of the order $(\hbar Qt)^{m+1}(\hbar^4t^3)^n$. For a given power
$k=4n-m-1$ of $\hbar$ their ratio to the leading contribution 
$(\hbar Qt)^k$ is again of order $(tQ^4)^{-n} \ll 1$. Finally,
for the terms of the kind (iii), their ratio to the leading terms
at a given power $k=4n$ is $(\hbar^4t^3)^n(\hbar Qt)^{-n} =
(tQ^4)^{-n} = (tQ^4)^{-k/4} \ll 1$. We thus conclude that each term
arising from the series of derivatives in (\ref{eq35}) is smaller
than the contribution from the leading term, and the relative
suppression increases with $k$ or $n$. However, it would be wrong
to conclude that these subdominant terms can all be neglected,
because there are certain terms among those of type (iii), which are 
independent of the cutoff $Q$ and which are not canceled by similar
contributions from the channels.

Having established that the terms arising from the logarithmic term 
in (\ref{eq35}) dominate in the expansion of $Z(t)$ in powers of $\hbar$, 
we now proceed to give those explicitly. The integrated form of the 
series (\ref{eq35a}) is:
\begin{equation}
\ln\coth\frac{u}{2} = - \ln\frac{u}{2} 
    + \sum_{k=1}^\infty \frac{(2^{2k-1}-1)}{k(2k)!} B_{2k} u^{2k} ,
\label{eq36b}
\end{equation}
This series converges for 
$u^2 < \pi^2$ or $(\hbar gtQ) < 2\pi$, which is consistent
with the condition permitting the use of the Wigner representation
in the square $[-Q,Q]$ and with the inequality $tQ^4 \gg 1$. 
We thus get for the partition function inside the four channels, up 
to terms involving powers of $(tQ^4)^{-1} \ll 1$:
\begin{eqnarray}
\label{eq36}
Z_{0+2+4+6+8}^{[Q,\infty]}(t) 
&=& K \left[ 4\ln\frac{4}{\hbar gtQ} + \frac{1}{12}(\hbar gtQ)^2 
      - \frac{7}{2^7\cdot 3\cdot 5!!} (\hbar gtQ)^4  \right.
\\
& & \qquad \left.
      + \frac{31}{2^9\cdot 3\cdot 9!!} (\hbar gtQ)^6 
      - \frac{127}{2^{16}\cdot 5\cdot 9!!} (\hbar gtQ)^8 + \cdots \right] .
\nonumber
\end{eqnarray}
We retained in (\ref{eq36}) the higher-order terms for future use.

The contributions $Z_0$ and $Z_2$ for the central square are given by 
(\ref{eq22}) and by eq.~(5) from \cite{bib7}. Their sum is:
\begin{equation}
Z_{0+2}^{[-Q,Q]}(t) = K \left[ 2\ln \sqrt{2}t^{1/2}g^2Q^2 + C - 
   \frac{1}{12}(\hbar gtQ)^2 \right] .
\label{eq37}
\end{equation}
Adding the first two terms in (\ref{eq36}) and (\ref{eq37}) we finally 
obtain for $Z$ up to the order of $\hbar^2$ the expression (\ref{eq04}) 
found in \cite{bib7}, 
i.~e.\ the $Q$-dependent corrections to the TF term vanish at the 
second-order of $\hbar$. Below we shall see that this statement is correct
even up to the order $\hbar^8$. We conclude that the unusually large number 
renormalizing the TF term in (\ref{eq27}) is an artefact of the neglect of 
the quantum character of the motion along the axis in the hyperbolic channels.

\section{The corrections to the TF term vanish up to order $\hbar^8$}

In this Section we show that the leading $Q$-dependent quantum corrections 
to the TF term up to the order $\hbar^8$ cancel if we treat the quantum 
mechanical motion in the channels correctly. 
Actually, we already gave the corrections to $Z(t$) up to $\hbar^8$ 
for the correct motion in the hyperbola channels (see formula (\ref{eq36})).
It remains for us to calculate these 
corrections for the square $x,y\in [-Q,Q]$, a straightforward though 
cumbersome task.   First of all,we need to know $W_4(t)$ using 
(\ref{eq10}).There are two types of contribution to $W_4(t)$: 
With $m-n=2$ and $m=n$. Terms with  $m-n=2$ give the main contribution  
of the order $tQ^4$, terms with $m=n$ contain only logarithms of 
$t Q^4$ and may be neglected  with the precision $\ln (tQ^4)/tQ^4\ll 1$. 
Here we give the expression for $W_4(t)$ after the integration over 
$p_x$ and $p_y$:
\begin{eqnarray}
\int d\Gamma\, W_4\, e^{-Vt} 
&=& \frac{2\pi g^2t^3}{2^4 t} \left( \frac{1}{5} g^2t I_{20} 
   -\frac{11}{45} (g^2t)^2 I_{31} \right.
\nonumber \\
& & \qquad\qquad
   +\frac{1}{36} (g^2t)^3 I_{42} - \frac{4}{15} I_{00} + g^2t I_{11}
\nonumber \\
& & \qquad\qquad \left.
   -\frac{17}{45} (g^2t)^2 I_{22} + \frac{1}{36} (g^2t)^3 I_{33}
   \right)
\label{eq41}
\end{eqnarray}
Integration of $I_{mn}$ over $x$ and $y$ for $m\neq n$ was already done 
before (see (\ref{eq22b})); for $m=n$ we obtain:
\begin{eqnarray}
I_{00} &=& \frac{\sqrt{2\pi}}{(g^2t)^{1/2}} 
           \left[\ln(g^2Q^4t) + C + \ln 2\right]
\nonumber \\
I_{11} &=& \frac{\sqrt{2\pi}}{(g^2t)^{3/2}} 
           \left[\ln(g^2Q^4t) + C + \ln 2 -2 \right]
\nonumber \\
I_{22} &=& \frac{3\sqrt{2\pi}}{(g^2t)^{5/2}} 
           \left[\ln(g^2Q^4t) + C + \ln 2 -\frac{8}{3} \right]
\nonumber \\
I_{33} &=& \frac{15\sqrt{2\pi}}{(g^2t)^{7/2}} 
           \left[\ln(g^2Q^4t) + C + \ln 2 -\frac{46}{15} \right] .
\label{eq41a}
\end{eqnarray}
Collecting all terms together we have from (\ref{eq41}) for the 
corrections to $Z(t)$ of the order of $\hbar^4$ in the square 
$x,y\in [-Q,+Q]$: 
\begin{eqnarray}
Z_4^{[-Q,Q]}(t) 
&=& K \left[ \frac{7}{2^7\cdot 3\cdot 5!!} g^4\hbar^4t^4Q^4 \right.
\nonumber \\
&& \qquad \left.
   + g^2\hbar^4t^3[\ln (g^2Q^2\sqrt{t/2}) - 8C - 16\ln 2 ] \right]
\label{eq41b}
\end{eqnarray}
where the first term comes from $I_{mn}$ with $m-n=2$, and the 
logarithmic terms, as we remarked before, arise from the contributions 
$I_{mm}$.  We see that the second term in (\ref{eq41b}) is of 
order $\ln (tQ^4)/Q^4t \ll 1$. Discarding it and adding the 
$\hbar^4$ correction from the channels in (\ref{eq36}) we find 
that with the precision $\ln (tQ^4 )/t Q^4 \ll 1$,
\begin{equation}
Z_4^{[-Q,+Q]} + Z_4^{[Q,\infty]} = 0  
\label{eq41c}
\end{equation}
as it was for the second-order corrections (with the precision 
$1/tQ^4$).  

Consider now the $\hbar^6$-order corrections to the partition
function. In (\ref{eq36}) we included the dominant contribution
from the channels $[Q,\infty]$ (fourth term). For $W_6$ in the
square we have, after integration over $p_x$ and $p_y$ and using
the notation of (\ref{eq21}):
\begin{eqnarray}
\int d\Gamma\, W_6\, e^{-tV} 
&=& \frac{2\pi (gt)^6}{t\cdot 9!!\cdot 2^6}
    \left[ - 61 I_{30} + \frac{249}{2}(g^2t) I_{41} \right.
\nonumber \\
&& \qquad \left. 
       - \frac{119}{4}(g^2t)^2 I_{52}
       + \frac{35}{24}(g^2t)^3 I_{63} \right] ,
\label{eq41d}
\end{eqnarray}
Using (\ref{eq22b}) for the evalulation of the $I_{mn}$ and collecting 
all terms, we obtain for $Z_6^{[-Q,Q]}$ in the
square:
\begin{equation}
Z_6^{[-Q,Q]}(t) = - K \frac{31}{3\cdot 2^9\cdot 9!!} (\hbar gtQ)^6 . 
\label{eq41f}
\end{equation}
Once more, the contribution from the square is exactly canceled
against the one from the four channels given in (\ref{eq36}),
as it happened for the second and fourth order:
\begin{equation}
Z_6^{[-Q,Q]}(t) + Z_6^{[Q,\infty]}(t) = 0
\label{eq41g}
\end{equation}
with the precision $\ln(tQ^4)/(tQ^4) \ll 1$.

Finally, we give here the order $\hbar^8$ corrections to $Z_8^{[-Q,Q]}$
in the square:
\begin{eqnarray}
\int d\Gamma\, W_8\, e^{-tV} 
&=& \frac{2\pi (gt)^8}{t\cdot 5!!\cdot 2^7}
\left[ \frac{1261}{7560} I_{40} 
     - \frac{259}{540} g^2t I_{51}
     + \frac{893}{5040} (g^2t)^2 I_{62} \right.
\nonumber \\
& & \qquad \left.
     - \frac{23}{1296} (g^2t)^3 I_{73}
     + \frac{5}{10368} (g^2t)^4 I_{84} \right] .
\label{eq41k}
\end{eqnarray}
where we again discarded terms proportional to $I_{mm}$, which are
of the order $\ln(tQ^4)/(tQ^4) \ll 1$ with respect to the terms
containing $I_{mn}$ with $m>n$. 
Using now the general expression (\ref{eq22b}) for $I_{mn}$ we obtain:
\begin{equation}
\int d\Gamma\, W_8\, e^{-Vt} = \frac{(2\pi)^{3/2} 127}
  {gt^{3/2}2^9\cdot 5!!\cdot 3\cdot 2^7\cdot 7!!} (gtQ)^8
\label{eq41l}
\end{equation}
and get the result
\begin{equation}
Z_8^{[-Q,Q]}(t) = K \frac{127}{5\cdot 2^{16}\cdot 9!!} (\hbar gtQ)^8
\label{eq41m}
\end{equation}
which miraculously cancels with the contribution from the channels
in (\ref{eq36}).

Although we cannot prove such a cancellation in general, to all orders,
we have no doubt that all higher-order corrections to the partition 
function containing powers of $(\hbar gtQ)$ cancel in the limit 
$tQ^4 \gg 1$. Of course, there are other corrections involving powers of 
$(tQ^4)^{-1}$, but these are suppressed due to the classical condition 
of adiabaticity $tQ^4 \gg 1$. We shall consider this issue in Section 7. 

In anticipation of later applications, we remark here that there is a
strong correlation between the power of $\hbar$, denoted by $k \geq 2$,
and power of the dominant terms in the limit $tQ^4 \gg 1$. A systematic
analysis of the higher-order corrections using {\em Mathematica} leads
to the conclusion that the difference between $m$ and $n$ in the leading
integral $I_{mn}$ from (\ref{eq21}) is $m-n=\frac{1}{2}k$. Besides the 
terms with the largest difference $m-n$, which give the leading contribution 
to $W_k$ and which we retain in (\ref{eq41d}) and (\ref{eq41k}), there are
also terms involving $I_{mn}$ with $m-n=\frac{1}{2}k-2\ell$ with
$\ell=1,2,\ldots <\frac{1}{4}k$. There is also a correlation between
the powers of $t$ and $m,n$. The analysis shows that for the terms 
with $m-n=\frac{1}{2}k$ the power of $t$ is $2m-n-1$, and for terms
with $m-n=\frac{1}{2}k-2\ell$ the power is $2m-n-1-3\ell$. The factor
of $g$ for $I_{mn}$ is $g^{2m}$. Straightforward calculations similar 
to the above show that the ratio of the contributions to the partition 
function for the terms $m-n=\frac{1}{2}k-2\ell$ to the ones for 
$m-n=k/2$ is of the order $1/(tQ^4)^\ell \ll 1$. Indeed, we
find:
\begin{eqnarray}
\label{eq41h}
Z_k^{(m,n)}(t) = K (\hbar gtQ)^k \frac{(2n-1)!!}{2^{n-1}}
\qquad \left( m-n=\frac{k}{2} \right) ;
\\ \nonumber
Z_k^{(m,n)}(t) = K (g^4tQ^4)^{-\ell} (\hbar gtQ)^k \frac{(2n-1)!!}
                 {2^{n-1}(k-4\ell)}
\qquad \left( m-n=\frac{k}{2}-2\ell \right) . 
\end{eqnarray}
This justifies neglecting these subdominant terms in our calculations 
here. 

Finally, at any power $k$ of $\hbar$ terms with logarithms ($m=n$) 
may be discarded with respect to the terms with powers of $Q^2$. Indeed, 
the last terms with $m>n$ are of the order of $(\hbar gtQ)^k$, terms with 
$m=n$ are of the order of $(g^2\hbar^4 t^3)^{k/4}\ln(tg^4Q^4)$ and may 
be neglected with the precision $\ln(g^4t Q^4)/(g^2tQ^4) \ll 1$. Again,
with increasing $k$ the precision improves.

\section{$Q$-independent terms in the channels and central region}

We already mentioned that the derivative expansion for the channel
contribution (\ref{eq35}) has one $Q$-independent term
for each power of of $\lambda^2=g^2\hbar^4t^3$. From the expansion
(\ref{eq35a}) it is evident that these terms have the form
\begin{equation}
\frac{2^{2n-1}-1}{n(2n)!}(2n-1)!! B_{2n} \lambda^{2n} .
\label{eq42}
\end{equation}
Each terms is of the order $(tQ^4)^{-1} \ll 1$ compared with the
leading term from the expansion of the logarithic term in (\ref{eq35}).
Collecting these terms we obtain a series of $Q$-independent terms
contributing to $Z(t)$:
\begin{equation}
\frac{1}{2^3\cdot 3}B_2\lambda^2 
  + \frac{2^3 -1}{2^{10}\cdot 3^2}B_4\lambda^4
  + \frac{2^5 -1}{2^{14}\cdot 3^5}B_6\lambda^6 
  + \frac{2^7 -1}{2^{23}\cdot 3^5}B_8\lambda^8 + \cdots
\label{eq43}
\end{equation}
We can rewrite this series in the form
\begin{equation}
\sum_{n=1}^{\infty} 
     \frac{2^{2n}(2^{2n-1}-1)(2n-1)!!}{2^{2(n-1)}n(2n)!} 
     B_{2n} \left(\frac{\lambda}{4\sqrt{3}}\right)^{2n} .
\label{eq44}
\end{equation}
The factor $(2n-1)!!$, which arises from the $(2n-1)$-th derivative 
in (\ref{eq35}), spoils the convergence of this series for $Z(t)$.
Using the identity $2^n n! (2n-1)!! = (2n)!$, we obtain for the
series of $Q$-independent terms:
\begin{equation}
Z^{[Q,\infty]} = K \sum_{n=1}^{\infty} 
     \frac{4(2^{2n-1}-1)}{2^n n! n} 
     B_{2n} \left(\frac{\lambda}{4\sqrt{3}}\right)^{2n} .
\label{eq45}
\end{equation}
This series is only asymptotic despite the smallness of $\lambda^2$,
because the Bernouilli numbers grow factorially at large $n$:
\begin{equation}
B_{2n} \to \frac{2(2n)!}{(2\pi)^{2n}} .
\label{eq46}
\end{equation}

Next we turn to the $Q$-independent contributions to $Z(t)$ from the
central region. We start with the investigation of the corrections
to the dominant terms. A straightforward calculation yields the
following correction factor to $I_{mn}$ from (\ref{eq22b}):
\begin{equation}
1 - \frac{(2m-1)!!}{(2n-1)!!} (g^2Q^4t)^{n-m} .
\label{eq47}
\end{equation}
This factor plays an important part making the corrected $I_{mn}$
well behaved in the limit $m=n$. Indeed, we note that with 
$\varepsilon = 2(m-n)$:
\begin{equation}
\frac{(2m-1)!!}{(2n-1)!!} 
\approx 1 + \varepsilon \sum_{\ell=1}^{m} \frac{1}{2\ell -1}
\label{eq48}
\end{equation}
the diagonal elements ($m=n$) become
\begin{equation}
I_{mm} = \lim_{\varepsilon\to 0} I_{mn} 
= \frac{\sqrt{2\pi}(2m-1)!!}{(g^2t)^{m+\frac{1}{2}}}
  \left[ \ln(g^2Q^4t) - 2 \sum_{\ell=1}^m \frac{1}{2\ell -1} \right] ,
\label{eq49}
\end{equation}
demonstrating that the corrections to $I_{mm}$ are independent of $Q$,
with the exception of the logarithmic term.
One easily confirms this general result by explicit calculations.
Substituting $x=w$ and $y=\sqrt{2/t}(u/gw)$ in (\ref{eq21}) we get:
\begin{equation}
I_{mm} = 4 \left( \frac{2}{g^2t} \right)^{\frac{2m+1}{2}}
  \int_0^Q \frac{dw}{w} \int_0^w du\, u^{2m} e^{-u^2} .
\label{eq50}
\end{equation}
We may write this expression as a derivative of the error function:
\begin{equation}
I_{mm} = 4 \left( \frac{2}{g^2t} \right)^{\frac{2m+1}{2}}
    \frac{\sqrt{\pi}}{2}(-1)^m \left. \frac{d^m}{da^m}
    \int_0^{w_0} \frac{dw}{\sqrt{a}w} {\rm erf}(\sqrt{a}w)
    \right|_{a=1} ,
\label{eq51}
\end{equation}
where $w_0^2 = g^2Q^4t/2$. Integrating by parts, discarding the 
exponentially small terms in $w_0$, then differentiating with respect
to $a$ and finally setting $a=1$, we obtain:
\begin{equation}
I_{mm} = \frac{\sqrt{2\pi}}{(g^2t)^{m+\frac{1}{2}}}
   \left[ (2m-1)!! \ln w_0^2 - \frac{2^m}{\sqrt{\pi}}
   \int_0^\infty dt\, t^{m-\frac{1}{2}} e^{-t} \ln t \right] .
\label{eq52}
\end{equation}
For the integral in the second term of this expression one finds
(see ref.~\cite{bib17} formula 4.352.2):
\begin{equation}
\int_0^\infty dt\, t^{m-\frac{1}{2}} e^{-t} \ln t
   = \frac{\sqrt{\pi}}{2^m} (2m-1)!! 
     \left[ 2 \sum_{\ell=1}^m \frac{1}{2\ell-1} - C - 2\ln 2 \right]
\label{eq53}
\end{equation}
which yields
\begin{equation}
I_{mm} = \frac{\sqrt{2\pi}(2m-1)!!}{(g^2t)^{m+\frac{1}{2}}}
   \left[ \ln(g^2Q^4t) + C + \ln 2 - 2 \sum_{\ell=1}^m \frac{1}{2\ell-1}
   \right] .
\label{eq54}
\end{equation}
Next we use a trick, first introduced by Euler, replacing the finite
sum by an asymptotic series (see e.g.~\cite{bib20}):
\begin{equation}
\sum_{\ell=1}^m \frac{1}{2\ell-1} = \frac{1}{2} \left[ C + \ln(2m)
   + \sum_{\ell=1}^\infty \frac{2^{2\ell -1}-1}{(8m^2)^\ell} B_{2\ell}
   \right] .
\label{eq55}
\end{equation}
We are now ready to combine the contribution from the central region
with the asymptotic series (\ref{eq45}) obtained earlier for the
$Q$-independent contribution for the channels. A special case is the
case $m=n=0$, for which there is no infinite sum, giving
\begin{equation}
I_{00} =  \sqrt{\frac{2\pi}{g^2t}}
          \left[ \ln(g^2Q^4t) + C + \ln 2 \right] .
\label{eq56}
\end{equation}
Taken together with the logarithmic term from the channels, this leads
to the $Q$-independent expression (\ref{eq04}) obtained in the improved
TF approximation \cite{bib7}. Our detailed analysis has shown that the
structures $I_{mm}$ containing $Q$-independent terms 
appear only in $W_{4k}$ with $k=1,2,\ldots$. This means that such terms 
appear only at the orders $\hbar^{4k}$ and implies that the expansion 
parameter of $Z(t)$ is $\lambda^2=g^2\hbar^4t^3$.

For a given power $\lambda^{2n}$ there are $3n$ quantities $I_{mm}$
($m=1,2,\ldots,3n$) and $I_{00}$, which we collect separately. The
results given above allow us to write the $Q$-independent asymptotic
series from the central region. Including also the TF term (\ref{eq04}),
we thus obtain:
\begin{eqnarray}
Z^{[-Q,Q]}(t) &=& K \left[ \ln\frac{1}{g^2\hbar^4t^3} + 9 \ln 2 + C \right.
\nonumber \\
  && + \sum_{n=1}^{\infty} \lambda^{2n} 
     \left( a_0^{(n)} (C+\ln 2) \right.
\nonumber \\ && \left. \left.
     - \sum_{m=1}^{3n} a_m^{(n)} (2m-1)!! \left[ \ln m 
     + \sum_{\ell=1}^\infty \frac{2^{2\ell -1}-1}{(8m^2)^\ell} B_{2\ell}
     \right] \right) \right] .
\label{eq57}
\end{eqnarray}
Here the $a_m^{(n)}$ ($m=0,1,2,\ldots,3n$) denote the numerical coefficients
in $(g^2t)^m I_{mm}$ occurring in $Z_{4n}$. For example, for $n=1$ we have 
four such numbers with alternating signs:
\begin{equation}
a_0^{(1)} = -\frac{1}{60} , \quad
a_1^{(1)} = \frac{1}{16} , \quad
a_2^{(1)} = -\frac{17}{720} , \quad
a_3^{(1)} = \frac{1}{576} .
\label{eq58}
\end{equation}

Summing up all these results, we may surmise that in the limit $tQ^4 \gg 1$
all $Q$-dependence is canceled, and we are left with two asymptotic series
(\ref{eq45}) and (\ref{eq57}) from the two regions contributing together
to the partition function of the two-dimensional YMQM model. We have shown
explicitly how the small parameter $(tQ^4)^{-1}$, artificially introduced
in the adiabatic separation of the degrees of freedom in the channels, 
transmutes into the genuine quantum mechanical parameter $\lambda^2 = 
g^2\hbar^4t^3$.

\section{Conclusions}

We have shown that the richness of the classical YM mechanics with
a $x^2y^2$ potential translates into, and even gets amplified by,
the quantum mechanical properties of the system. The YM quantum 
mechanics exhibits a confinement property, which strongly influences
the quantum mechanical motion in the $x^2y^2$ potential. At higher
order in $\hbar$ (up to $\hbar^8$) this results in the vanishing of 
the leading quantum corrections (for $tQ^4 \gg 1$), when we correctly 
take into account this property for the motion in the hyperbolic 
channels. We believe that this result may survive to even higher order.
We also derived a novel form of the equation for the Uhlenbeck-Beth
function $W({\vec r},{\vec p};t)$, which is the basis of the WK
expansion. We believe that our expression can be a starting point 
for new approximation schemes using techniques from diffusion theory.
We hope that the lessons derived from the present study of the
higher-order quantum corrections to the homogeneous limit of the 
Yang-Mills equations will be useful for an improved understanding 
of the internal dynamics of the Yang-Mills quantum field theory.

\end{document}